\documentclass[twocolumn,showpacs,preprintnumbers,amsmath,amssymb]{revtex4}
\usepackage{epsfig}
\usepackage{graphicx}% Include figure files
\usepackage{dcolumn}% Align table columns on decimal point
\usepackage{bm}% bold math

\begin{document}

% \preprint{preprint}

\title{Colloidal glass transition observed in confinement}

\author{Carolyn R.~Nugent}
\altaffiliation{Current address:
Geophysics and Space Physics, UCLA, Los Angeles, CA}
%\affiliation{Physics Department, Bucknell University, Lewisberg, PA}
\author{Kazem V.~Edmond}%
\author{Hetal N.~Patel}%
\author{Eric R.~Weeks}%
 \email{weeks@physics.emory.edu}
\affiliation{Physics Department, Emory University, Atlanta, GA 30322}

\date{\today}% It is always \today, today,

\begin{abstract}
We study a colloidal suspension confined between two
quasi-parallel walls as a model system for glass transitions in
confined geometries.  The suspension is a mixture of two
particle sizes to prevent wall-induced crystallization.
We use confocal microscopy to directly observe the motion of
colloidal particles.  This motion is slower in confinement,
thus producing glassy behavior in a sample which is a liquid
in an unconfined geometry.  For higher volume fraction samples
(closer to the glass transition), the onset of confinement
effects occurs at larger length scales.
\end{abstract}%  86 words today (5/3/07)

% glass transition, glasses, colloids

\maketitle

% ===================================================
% \section{Introduction}
% ===================================================

Glasses are typically formed by rapidly quenching the
temperature of a liquid, resulting in an amorphous
liquid-like microstructure with macroscopic solid-like
behavior.  Upon approaching the glass transition, the
temperature might be changed by only a factor of two while
simultaneously the viscosity of the liquid grows by many
orders of magnitude \cite{reviews}.  A conceptual microscopic
explanation for the viscosity growth is the idea of dynamic
length scales: in order for molecules in the material to
rearrange, they must move together as a group.  As the glass
transition is approached, the increasing 
size of these groups relates to the increasing macroscopic viscosity
\cite{reviews,mckenna05,ediger00,weeks00,kegel00,weeks07}.

An important way to probe these length scales is to
study the behavior of glass-forming systems when they
are confined, to constrict the range of accessible length scales
\cite{mckenna05,kob00,lowen99,robbins92,yamamoto00,jackson91,barut98,richert94,morineau02}.
Intriguingly, the glass transition temperature $T_g$
can be both smaller or larger in confined geometries
\cite{jackson91,barut98,richert94}, even for the same
material \cite{mckenna05,kob00,morineau02}.
Experiments and simulations suggest that
the interaction between the confining surface and the sample
is crucial.  For strong interactions (or atomically rough
surfaces) the glass transition happens ``sooner,'' that is,
confinement increases $T_g$ by slowing motion near the surfaces
\cite{mckenna05,kob00,lowen99,richert94,morineau02}.
Likewise, for systems that weakly interact with the walls,
$T_g$ is typically smaller
\cite{mckenna05,kob00,jackson91}.  However, a clear explanation
of these phenomena is still lacking.  As it is difficult to
get details out of experiments \cite{mckenna05}, the use
of computer simulations to visualize the motion is important 
\cite{kob00,lowen99,yamamoto00,robbins92}.

We use confocal microscopy to directly visualize the motion
of colloidal particles, which serve as a model system for
the glass transition in confinement.  Colloids undergo a
glass transition in bulk samples as the solid particle volume
fraction $\phi$ is increased
\cite{weeks00,kegel00,pusey86,vanmegen96}.
At high volume fraction near the colloidal glass
transition ($\phi_g \approx 0.58$), particles
move in rearranging groups characterized by a length scale
of $\sim$3-6 particle diameters \cite{weeks00,weeks07}, similar to
simulations \cite{yamamoto00}.  In this manuscript we study a
mixture of two sizes of colloidal particles confined between two
quasi-parallel plates, with a plate gap as small
as 3.0 large-particle diameters.  In our system confinement
induces the glass transition ``sooner,'' at concentrations for
which the bulk behavior is still liquid-like.  Studying the
glass transition in confinement may help us understand the
glass transition in the bulk \cite{mckenna05}.  Furthermore,
understanding the properties of confined liquids has relevance
for lubrication \cite{robbins92,granick99}, dusty plasmas
\cite{teng03}, and the flow of glassy complex fluids through
microfluidic devices \cite{granick99}.

% ===================================================
% \section{Experimental Methods}
% ===================================================

Our colloidal samples are poly-methyl(methacrylate)
particles, sterically stabilized to prevent aggregation
\cite{weeks00,pusey86}.  We use a mixture of two particle
sizes, with radii $a_{\rm small} = 1.18$ $\mu$m and $a_{\rm
large} = 1.55$ $\mu$m.  While the particle polydispersity is
low ($\sim$5\%), the mean particle radii are only known to
within $\pm 0.02$ $\mu$m.  The mixture of two particle sizes
prevents crystallization which would otherwise be induced by
the walls \cite{lowen99,teng03,murray98}.  In each sample,
the small particles are dyed with rhodamine dye, and the large
particles are undyed.  We use a mixture of cyclohexylbromide
and decalin as our solvent, to match the density and index
of refraction of the particles; the particles are slightly
charged in this solvent \cite{dinsmore01}.  The viscosity of
the solvent is 2.25 mPa$\cdot$s.  We examine four different
samples A-D, with properties listed in Table I.

% TABLE ONE
\begin{table}
\caption{\label{table1}
Characteristics of the four samples studied.
The number ratio $N_{\rm small}/N_{\rm large}$ is determined
by counting particles in several fields of view using DIC
(differential interference contrast) microscopy.
The total volume fraction $\phi_{tot}$ is determined using
confocal microscopy, by counting
the number of small particles seen in a given imaging volume,
using the known number ratio to determine the number of large
particles present, and then using the particle sizes and the
imaging volume size to compute $\phi_{tot}$.  Additionally
$\phi_{tot}$ and $N_{\rm small}/N_{\rm large}$ was confirmed in
samples B--D by direct 3D confocal microscopy observation, where
the particle sizes could be easily distinguished and counted;
the results were in agreement with the DIC measurements.
The volume fractions of the small species and large species,
$\phi_s$ and $\phi_l$, are calculated from the other two
quantities.
The uncertainties of $N_{\rm small}/N_{\rm large}$ are $\pm
5\%$, and the uncertainties of $\phi_{tot}$ are $\pm 8\%$.
In particular, note that samples A and C likely do not have
the same volume fraction, but it is unclear which has the larger
$\phi_{tot}$.  Samples B, C, and D are prepared by dilutions
of one stock sample and thus all have the same $N_{\rm
small}/N_{\rm large}$.
}
\begin{ruledtabular}
\begin{tabular}{ccccc}
Sample & $N_{\rm small}/N_{\rm large}$ & $\phi_s$ & $\phi_l$ &
$\phi_{tot}$\\
\hline
A & 3.5 & 0.26 & 0.16 & 0.42 \\
B & 3.0 & 0.13 & 0.10 & 0.23 \\
C & 3.0 & 0.24 & 0.18 & 0.42 \\
D & 3.0 & 0.26 & 0.20 & 0.46 \\
\end{tabular}
\end{ruledtabular}
\end{table}

% Initial conditions:  try to disturb the slide as little as
% possible (see frnugent.060105a).

% We use DIC microscopy to verify that both species of particles
% move freely.

We observe our samples using confocal microscopy
\cite{dinsmore01,prasad07}.  As the larger particles are
not dyed, we only see the smaller particles.  We use a fast
confocal microscope (VT-Eye from Visitech, International)
with a 63$\times$ air objective (N.A.~0.70) to scan a volume
$50 \times 50 \times 20$ $\mu$m$^3$ once every 2.0~s over a
period of an hour.  We analyze the images offline to locate the
positions of visible (smaller) particles, with a resolution
of 0.05~$\mu$m in $x$ and $y$ (parallel to the walls) and a
resolution of 0.1~$\mu$m in $z$ (parallel to the optical axis).
We then track their motion in 3D \cite{dinsmore01}.

Our sample chambers are made by placing a microscope
coverslip at a slight angle, supported by a small piece of
mylar film (thickness 100 $\mu$m) at one end and resting
directly on the microscope slide at the other end
\cite{murray98,palberg05}.  The ends and sides are
sealed with UV-curing epoxy.  Thus a thin wedge-shaped chamber
is formed with an angle $\approx 0.4^\circ$, ensuring that
locally the walls are essentially parallel, and allowing us
to study a single sample at a variety of different confinement
thicknesses \cite{murray98,palberg05}.

The glass surfaces of the coverslip and slide are untreated.
In experiments with sample A, we find that some colloidal
particles stick to these surfaces.  
The stuck particle coverage 
is typically 10\% - 20\% of the area.
In a second series
of experiments done with samples B--D, no particles were stuck.
Reassuringly, we find little dependence of
the behavior on the number of stuck particles in the results
discussed below \cite{stuck}.

%% This is easily deduced from the data: some particles adjacent
%% to walls do not move, while the others move freely, with a
%% clear distinction between the two types.

For sample A, measuring the positions of the stuck particles
allows us to accurately measure the sample thickness.
While the uncertainty in locating individual particle positions
in $z$ is 0.1~$\mu$m, by averaging data from tens of
stuck particles over hundreds of images we locate their mean
$z$ position to better than 0.005~$\mu$m.  Thus the effective
thickness $H$ of each experimental data set is determined to
within 0.01~$\mu$m, and is the range in $z$ available to the
{\em centers} of the visible particles.  In this manuscript
our thicknesses are reported in terms of $H$.  The true
surface-to-surface thickness of a sample chamber is found by
adding $2a_{\rm small} = 2.36$~$\mu$m to $H$.

% The uncertainty of $a_{\rm small}$ implies a systematic
% uncertainty for the true surface-to-surface thickness.

%   The larger particles with radius $a_{\rm large}$ have a slightly
%   smaller range in $z$ available to them, that is, $H + a_{\rm
%   small} - a_{\rm large}$.

% ===================================================
% \section{Results}
% ===================================================

\begin{figure}
\centerline{
\epsfxsize=8.0truecm
\epsffile{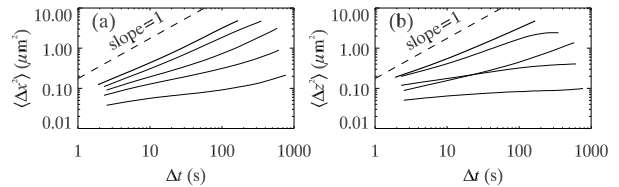}}
% \smallskip
\caption{Mean square displacements.  (a) Data for sample A,
showing motion parallel to the walls, for thicknesses $H$ =
bulk, 16.28 $\mu$m, 11.06 $\mu$m, 9.41 $\mu$m, and 6.92 $\mu$m
(from top to bottom).  The dashed line has a slope of 1 and
indicates the expected motion for a very dilute bulk suspension
of particles with radius $a_{\rm small}$.
(b) Similar to (a), but for motion perpendicular to the walls.
Data are ordered by thickness as $\Delta t \rightarrow \infty$,
as in (a).
}
\label{msd}
\end{figure}

For the first series of experiments, we study the behavior of
sample A ($\phi \approx 0.42$) as a function of thickness.
We quantify the particle motion
by calculating the mean square displacement (MSD),
$\langle \Delta x^2 \rangle = \langle (x_i(t+\Delta t) -
x_i(t))^2 \rangle$,
where the average is taken over all particles $i$ and all
initial times $t$, and a similar formula applies for $\langle
\Delta y^2 \rangle$ and $\langle \Delta z^2 \rangle$.  We find
that $\langle \Delta x^2 \rangle \approx \langle \Delta y^2
\rangle$ for all our experiments; we report our results
for the $x$ direction, the direction over which the
sample chamber has constant thickness.  We first consider the
results for motion parallel to the confining plates, $\langle
\Delta x^2 \rangle$, shown in Fig.~\ref{msd}(a).  The upper
bold line shows motion in an unconfined
region and is reproducible for all chamber thicknesses $H >
20$~$\mu$m.  For this sample, the motion in the unconfined
region is nearly diffusive, with the MSD growing almost
with slope 1 on the log-log plot.
This behavior is similar to monodisperse samples with a volume
fraction of $\phi \leq 0.4$ \cite{weeks00}.  In other words,
this sample is far from the glass transition, $\phi_g \sim 0.6$
\cite{sciortino04,vanmegen96}.
In thinner regions, the motion slows, as seen in the
sequence of solid curves below the top bold curve in
Fig.~\ref{msd}(a).  This slowing starts at a thickness of
$H \sim 16$~$\mu$m (2nd curve from top) and slows dramatically
for thinner samples; note that Fig.~\ref{msd}(a) shows a
log-log plot and thus for the thinnest region shown (bottom
curve, $H=6.92$~$\mu$m), to move a distance $\langle x^2 \rangle =
(a_{\rm small}/3)^2$ it takes a time scale 180 times larger
than for the bulk region data ($\Delta t = 500$~s as compared to 2.8 s).

%% We take three-dimensional data to resolve the 3D particle
%% motion between the two confining walls.

%% , although not as rapidly
%% as a dilute suspension [dashed line in Fig.~\ref{msd}(a)].

% FIG - MSD - ONE
% part A,B:  112c, 118e, 1189c, 119d, 1111cb
%For all curves, the small particle number density $n_{\rm
%small}$ is within 30\% of the bulk region value.  

%% (c) Data for samples B--D with volume fractions as indicated,
%% for thicknesses $H$ = bulk (solid lines) and $H = 20$~$\mu$m
%% (circles).  Data shown is motion parallel to the walls.  (d)
%% Data for densest sample (sample D, $\phi=46$) for thicknesses
%% 54 $\mu$m, 43 $\mu$m, 38 $\mu$m, 36.5 $\mu$m, 35 $\mu$m,
%% 33 $\mu$m, 25 $\mu$m, and 18 $\mu$m (from top to bottom).

%   at a time scale of $\Delta t =
%  100$~s, the value of the MSD is reduced by a factor of 40.

These results suggest that confinement induces glassy behavior,
with the influence of confinement beginning at $H \approx
16$~$\mu$m~$\approx 14 a_{\rm small} \approx 10 a_{\rm large}$
for this sample.  For the lower curves in Fig.~\ref{msd}(a),
the characteristic behavior of a ``super-cooled'' sample is
seen:  at shorter lag times ($\Delta t < 100$~s),
the MSD has a plateau, while at longer lag times, the MSD begins to rise
again \cite{kob00,weeks00}.  (For short time scales, particles
are trapped in cages formed by their neighbors, causing the
plateau in $\langle x^2 \rangle$.  At longer time
scales, these cages rearrange \cite{weeks00,weeks07}.)  For the thinnest region
(bottom curve), the particles remain localized for the duration
of the experiment.

%% , with only a hint of motion at long lag times.

% In thin regions, this sample behaves
% like a bulk sample with a higher volume fraction.

The slowing is also seen in motion perpendicular to
the walls, quantified by $\langle \Delta z^2 \rangle$,
shown in Fig.~\ref{msd}(b).  Moreover, in comparison with
Fig.~\ref{msd}(a), it is seen that the motion perpendicular to
the walls, $\langle \Delta z^2 \rangle$, is slowed even more so
than motion parallel to the walls, $\langle \Delta x^2 \rangle$.
This is not surprising, given that particles close to the walls
cannot move toward the walls at all, whereas motion parallel
to the walls is less restricted.

\begin{figure}
\centerline{
\epsfxsize=8.0truecm
\epsffile{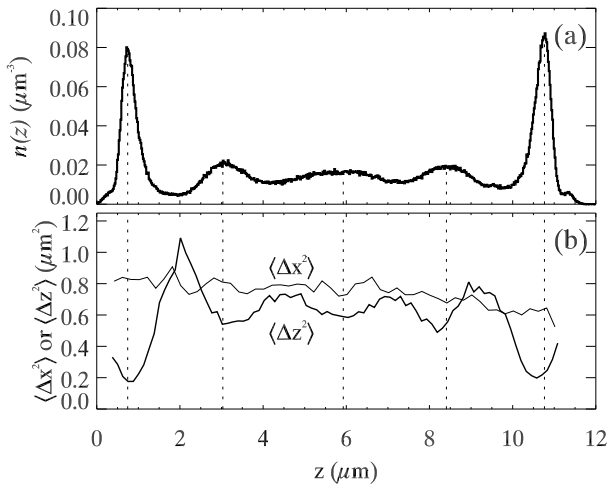}}
% \smallskip
\caption{
(a) Particle number density $n_{\rm small}(z)$ as a function
of distance $z$ across the sample cell.  Additional particles
are stuck to the walls of the sample cell (not shown in
the plot) which have centers located at $z=0.00$ $\mu$m
and $z=H=11.06$~$\mu$m.  This data corresponds to the middle
curve in Fig.~\ref{msd}(a), that is, a sample with moderately
slowed dynamics.
(b) Mean square displacement parallel to the walls ($x$) and
perpendicular to the walls ($z$), as a function of $z$.
The displacements are calculated using $\Delta t = 100$~s.  The vertical dotted
lines indicate the positions of the peaks
from part (a).  Data shown are from sample A ($\phi \approx 0.42$).
}
\label{dz}
\end{figure}
% data from 1189c

More than merely constricting motion, the walls also induce a
layering of particles, as seen in Fig.~\ref{dz}(a), similar to
simulations \cite{lowen99,archer07}.  The layering is most pronounced
immediately adjacent to the walls.  The centers of these
peaks are not at the precise distance $a_{\rm small}$ from
the walls, but are slightly offset toward the interior of the
sample.  (The centers of the stuck particles indicate
the maximum possible extent in $z$ that particles could be
located, and correspond to the ``feet'' of the data shown in
Fig.~\ref{dz}(a) at $z=0$ and 11.06~$\mu$m.  These particles
are not counted in $n_{\rm small}$ shown in Fig.~\ref{dz}(a).)

% As the sample is a mixture of two particle sizes, the layers
% farther from the walls are less pronounced than what would be
% seen in a monodisperse sample \cite{lowen99}.

The layers influence the dynamics, as seen in Fig.~\ref{dz}(b),
which shows how $\langle \Delta x^2 \rangle$ and $\langle
\Delta z^2 \rangle$ depend on $z$.  The displacements are
calculated using $\Delta t = 100$~s, as a representative time
scale over which particles begin to move out of this cage,
although the results do not depend on this choice and are similar
for caged behavior ($\Delta t = 2.0$~s for example).
Particles in the layers [the peaks of $n(z)$] have smaller
vertical displacements, as seen by the dips in $\langle \Delta
z^2 \rangle$ (heavy line).  The implication is that particles
in layers are in a preferred structure and less likely to
move elsewhere \cite{lowen99,archer07}.

%% Figure \ref{dz}(b) shows that the overall value of $\langle
%% \Delta x^2 \rangle$ (thin line), while higher than $\langle
%% \Delta z^2 \rangle$ (thick line), is reduced from the
%% behavior in the bulk region.  For this $\Delta t = 100$~s,
%% the bulk region data have $\langle \Delta x^2 \rangle \approx
%% 2$~$\mu$m$^2$/s, as seen in Fig.~\ref{msd}(a).

Surprisingly, the layers do not appear to influence the
motion parallel to the walls, as seen by $\langle \Delta
x^2 \rangle$ (thin line), which does not depend on $z$.
(The slight dip in $\langle \Delta x^2 \rangle$ seen at the
largest values of $z$ is not reproducible in other data sets.)
This seems counterintuitive as hydrodynamic interactions
with the wall normally result in reduced motion for nearby
particles \cite{dufresne00}.  We speculate that the cage
dynamics dominate particle motion, rather than hydrodynamic
influences \cite{squires}.  For example, if a particle is pulled by an
external force in a direction parallel to the walls, other
particles would be forced to rearrange, which is probably
the most significant contribution to the drag.
Particle rearrangements would be even more constrained for a
particle pulled perpendicular to the wall, thus explaining why
we observe slower $z$ motion \cite{squires}.
Simply put, the high volume fraction likely results in
hydrodynamic screening.

Thus while confinement causes the layering of particles near
the walls, this layering does {\em not} appear directly
responsible for the slowing of the particle motion.
Rather, the layering seems to be an additional influence on
the motion in the direction perpendicular to the walls, as
seen in Fig.~\ref{dz}(b), but only a minor influence compared
to the overall fact of confinement.  Note that 
results do not appear to depend on having an integral
number of well-defined layers between the walls \cite{robbins92}.  
The overall dynamics slow smoothly and monotonically as
the confining dimension decreases.

Our observation that the layers closest to the wall have
slower motion perpendicular to the walls agrees qualitatively
with previous experiments \cite{barut98,richert94,morineau02}
and simulations \cite{kob00} which suggested that surface
layers may be glassier than the interior.  However, we note
in our experiment this is strongly directionally dependent.
The slowing is most easily seen if $\langle \Delta z^2 \rangle$
can be measured independently of the other two directions.

%% Measurements of $\langle \Delta r^2 \rangle = \langle \Delta
%% x^2 \rangle + \langle \Delta y^2 \rangle + \langle \Delta z^2
%% \rangle$ only show a slight slowing near the walls, given that
%% the $x$ and $y$ directions aren't slowed [Fig.~\ref{dz}(b)].

%% We study the behavior of the wall layers in comparison
%% with the interior of the sample in Fig.~\ref{wall},
%% which shows the mean square displacement calculated
%% from different populations of particles.  The motion of
%% particles in the interior ($2.0$~$\mu$m$<z<10.0$~$\mu$m) is
%% nearly the same in $x$ and $z$, as indicated by the lines
%% in Fig.~\ref{wall}.  For particles in the surface layers
%% ($z<2$~$\mu$m or $z>10$~$\mu$m), motion parallel to the walls
%% ($\langle \Delta x^2 \rangle$) is unchanged (open circles).
%% However, motion perpendicular to the walls is greatly slowed
%% (filled circles); these particles behave even glassier than
%% the particles in the interior \cite{richert94,morineau02}.
%% However, this slowing at the walls is less pronounced than
%% what has been seen in simulation \cite{kob00}.

% .  These particles require a much larger time scale $\Delta t$
% to move any given distance $\langle \Delta z^2 \rangle$.  Thus

\begin{figure}
\centerline{
\epsfxsize=8.0truecm
\epsffile{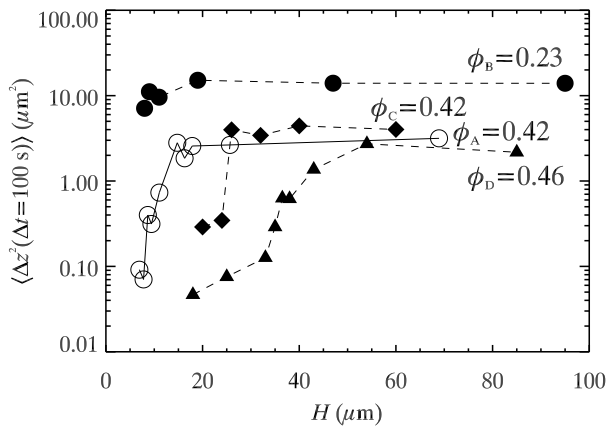}}
% \smallskip
\caption{
Value of $\langle \Delta x^2 \rangle$ at $\Delta t = 100$~s, as a
function of thickness $H$, for samples with $\phi$ as indicated.
The open circles correspond to sample A with 
$N_{\rm small}/N_{\rm large}=3.5$,
while the solid symbols correspond to samples B--D with 
$N_{\rm small}/N_{\rm large}=3.0$.
The lines are drawn to guide the eye.  The
plateau for each data set indicates behavior corresponding to the
bulk, whereas the downturn at low $H$ gives an idea of the length
scale at which confinement becomes important.
}
\label{xh}
\end{figure}

As noted earlier, the growth of dynamic length scales has been
observed as the glass transition is approached in a bulk material
\cite{reviews,mckenna05,ediger00,weeks00,kegel00,weeks07}.
For our colloidal samples, this implies that samples with a
larger $\phi$ should exhibit stronger confinement effects.
To check this, we took data from samples B, C, and D at
various thickness.  Qualitatively the data resemble that
shown in Fig.~\ref{msd}(a).  To capture the $H$ dependence,
Fig.~\ref{xh} shows the values of $\langle \Delta x^2 \rangle$,
at fixed $\Delta t = 100$~s, as a function of $H$ for the
different samples.  Consider the solid triangles, corresponding
to sample D.  For $H > 50$~$\mu$m, $\langle \Delta x^2 \rangle$
is essentially constant.  At $H <
50$~$\mu$m, the data start showing a strong $H$ dependence,
suggesting a confinement length scale of $H^* \approx
50$~$\mu$m.  For the solid symbols, an increase in $H^*$
is seen as $\phi$ increases, from approximately 10~$\mu$m
to 50~$\mu$m, confirming that there is a growing length
scale as the glass transition is approached.  These length
scales are significantly larger than those seen for dynamical
heterogeneities in monodisperse samples, which are 4 -- 8
$\mu$m \cite{weeks07}.  However, this agrees with simulations
which found a confinement length scale significantly larger
than the mobile cluster size \cite{kob00}.  In Fig.~\ref{xh},
sample A has a smaller value of $H^*$ relative to sample C,
which may be due to the excess of small particles in sample A;
see Table I.

%% In Fig.~\ref{msd}(c), we plot the mean square displacement from
%% samples B, C, and D, taken from the bulk region (lines) and from
%% $H = 20$~$\mu$m (circles).  As the volume fraction is increased
%% from $\phi_B=0.23$ to $\phi_D=0.46$, the behavior in the bulk
%% region slightly slows down.  For sample B, confined motion
%% at $H=20$~$\mu$m coincides exactly with the unconfined data.
%% For sample C,

% FIG - XH - THREE

% In the bulk region, sample A is at too low a volume fraction
% for the particles to rearrange in groups (using the methods of
% \cite{weeks00} to look for such groups) -- any length scale for
% such groups appears to be at most $\sim 2$ particle diameters.
% It is intriguing that we observe the confinement effects to
% begin at a thickness $H$ of approximately 7 small particle
% diameters, which suggests a new length scale.  

We find that confinement slows the motion of colloidal particles
and thus induces a glass transition to occur sooner than normal,
in other words, at volume fractions for which the bulk behavior
is liquid-like.  Simulations suggest the roughness of the walls
is crucial to this slowing \cite{kob00,lowen99} and we plan to
vary this in future experiments.  However, we note that our data
show slowing both with completely smooth walls (samples B,
C, and D) and walls with isolated stuck particles (sample A).
In contrast to our work, rough walls in simulations are composed
of particles fixed in a liquid-like structure
\cite{kob00,lowen99}.  This prevents layering of adjacent particles
and restricts motion parallel to the walls.  Thus the glass
transition in confined samples occurs sooner (at higher
temperatures \cite{kob00} or lower densities \cite{lowen99}).
In our experiments, particle motion parallel to the wall is
not noticeably inhibited, as seen in Fig.~\ref{dz}.  Yet,
we still find the glassy behavior occurs sooner:  at constant
volume fraction, the dynamics are slower as the confining
dimension decreases.  Thus it seems that the important effect
in our experiments is simply the restriction of motion perpendicular
to the wall, close to the surface of the wall.

% prevention of layering is not necessary to induce
% the glass transition \cite{layers}, nor is the restriction of
% motion parallel to the wall crucial.  Rather, the 

We thank A.~Schofield and W.~C.~K.~Poon for our colloidal
samples, and J.-L.~Barrat, E.~R.~Dufresne, M.~F.~Hsu,
W.~Kob, J.~Saldana, and T.~M.~Squires for helpful discussions.
Initial support for this project was provided by the Donors
of The Petroleum Research Fund, administered by the American
Chemical Society (37712-G7).  The work of CRN, KVE, and ERW
was also supported by the National Science
Foundation under Grant No.~DMR-0239109.


\begin{thebibliography}{99}
\bibitem{stuck} We conjecture that the sticking process occurs
over a long time period.  A year passed between making the
first sample chamber and the data acquisition, whereas the
second series of samples were investigated within a few weeks
of making the sample chambers.
\end{thebibliography}
\end{document}